\documentclass[preprint,preprintnumbers,amsmath,amssymb]{revtex4}
\usepackage{graphics,epsfig,subfigure}
\usepackage{graphicx}
\usepackage{color}
\usepackage{multirow}
\usepackage{array}
\usepackage{ulem}
\usepackage{indentfirst}
\usepackage{fancyref}
\usepackage{hyperref}
\usepackage{epstopdf}
\hypersetup{hypertex=true,
           colorlinks=true,
            linkcolor=blue,
            anchorcolor=blue,
            citecolor=blue}

\newcommand\beq{\begin{equation}}
\newcommand\eeq{\end{equation}}
\newcommand\beqn{\begin{eqnarray}}
\newcommand\eeqn{\end{eqnarray}}

\begin{document}

\title{Overcharging a Reissner-Nordstr\"om Taub-NUT regular black hole}

\author{Wen-Bin Feng$^{a,b}$,
	Si-Jiang Yang$^{a,b}$,
	Qin Tan$^{a,b}$,
	Jie Yang$^{a,b}$,
	and Yu-Xiao Liu$^{a,b}$\footnote{liuyx@lzu.edu.cn, corresponding author}}
\affiliation{$^{a}$Lanzhou Center for Theoretical Physics, Key Laboratory of Theoretical Physics of Gansu Province, School of Physical Science and Technology, Lanzhou University, Lanzhou 730000, China\\
    $^{b}$Institute of Theoretical Physics $ \&$ Research Center of Gravitation, Lanzhou University, Lanzhou 730000, China}

\begin{abstract}

The destruction of a regular black hole event horizon might provide us the possibility to access regions inside black hole event horizon. This paper investigates the possibility of overcharging a charged Taub-NUT regular black hole via the scattering of a charged field and the absorption of a charged particle. For the charged scalar field scattering, both the near-extremal and extremal charged Taub-NUT regular black holes cannot be overcharged. For the test charged particle absorption, the result shows that the event horizon of the extremal charged Taub-NUT regular black hole still exists while the event horizon of the near-extremal one can be destroyed. However, if the charge and energy cross the event horizon in a continuous path, the near-extremal charged Taub-NUT regular black hole might not be overcharged.
\end{abstract}

\maketitle

\section{Introduction}

The discovery of gravitational waves \cite{LIGO16,LIGO17,LIGO20}, generated by black hole or neutron star binaries, opened a new window towards highly relativistic systems, such as cosmology \cite{Zhan19} and black holes \cite{LCHu19}. Recently, gravitational wave has become a powerful tool to help us to distinguish general relativity from other gravity theories \cite{CFMPR17,BDFM18,FLLZC19,Heis18,GBPU16,NiZh19,Isha18,FSCMMF17,HJCa19}. Furthermore, gravitational waves aroused a lot of interest to study properties of black holes, such as no hair theory \cite{IGFST19}, quasi-topological properties \cite{LMLL20}, and thermodynamics \cite{KuMa19,WuWu19,MiXv19,BoGK19,ChJi19,WWYu19}. However, the existence of spacetime singularities might indicate the failure of predictability of gravity theory. To protect the predictability of gravity theory, Penrose proposed the weak cosmic censorship conjecture, which states that spacetime singularities are always hidden inside the event horizons of black holes and cannot be seen by distant observers \cite{Penr69}. Although the conjecture is still unproved, it has become one of the foundations of black hole physics. The existence of event horizons protects the predictability outside the event horizons. However, it also makes {it} hard to access the black hole interior by experiment. Thus, the destruction of black hole event horizons would provide the possibility of observing regions inside black holes and relevant new physics to build a consistent theory of quantum gravity.

There are many ways to consider the violation of the weak cosmic censorship conjecture \cite{East19,FiKT16,FKLT17,CrSa17,HSSLW20,SoHW20}. The seminal work that tried to destroy the event horizon of an extremal black hole in order to violate the weak cosmic censorship conjecture was first proposed by Wald, where a test particle with large charge or large angular momentum was thrown into an extremal Kerr-Newman black hole. The result suggests that particles causing the destruction of an event horizon would not be captured by the black hole \cite{Wald74}. The systematical works of Rocha and Cardoso et al. for the BTZ black hole \cite{RoCa11}, higher-dimensional Myers-Perry family of rotating black holes and a large class of five-dimensional black rings \cite{LCNR10,ShDA19} support the result that extremal black holes cannot be destroyed in the test particle approximation. While, Hubeny extended the research to near-extremal black holes and found that a near-extremal charged black hole can ``jump over" the extremal limit and become a naked singularity \cite{Hube99}. Further research of Jacobson and Sotiriou suggests that a near-extremal Kerr black hole can be overspun \cite{JaSo09}. Recently, Li and Bambi considered the destruction of the event horizons of regular black holes, such as Bardeen and Hayward black holes. They demonstrated that the event horizons of these black holes can be destroyed by test particles, and claimed that the destruction of such event horizons might provide us the possibility to access regions inside black hole event horizons \cite{LiBa13}. However, using the new version of gedanken experiment proposed by Sorce and Wald \cite{SoWa17}, when the second-order approximation of the perturbation that comes from the matter fields {was} taken into account, Jiang and Gao showed that a static charged regular black hole coupled to nonlinear electromagnetical field cannot be overcharged \cite{JiGa20}, and {much work has} been done to destroy the event horizons of black holes using this new version of gedanken experiment \cite{QYWR20,ZhJi20,JiZh20,Jian20}.

Another intriguing method of destroying the event horizon of a black hole is by using the scattering of a test classical or quantum field. The scattering of a classical field to destroy the event horizon was first proposed by Semiz and the result shows that the event horizon of an extremal dyonic Kerr-Newman black hole cannot be destroyed \cite{Semi11}. This method was further developed by others \cite{DuSe13,SeDu15,Duzt15}. Recently, Gwak considered that the change of a Kerr-(anti) de Sitter black hole is infinitesimal since the scattering process happens during the infinitesimal time interval, and found that the black hole cannot be overspun \cite{Gwak18}.  More studies {following} this line can be found in Refs. \cite{Gwak19a,Gwak19b,Chen18}. This consideration of the scattering process in an infinitesimal time interval may imply that the time interval for particles across the event horizon may play an important role {in destroying} the event horizon \cite{YCWWL20,YWCYW20,LiWL19,Gwak17}. Furthermore, when quantum mechanics is taken into account, near-extremal black holes may indeed be destroyed by the scattering of quantized fields \cite{MaSi07,Hod08,RiSa08,MRSSV09,RiSa11}.

Recently, black holes with NUT {parameters} have been investigated in many aspects, such as gravitational lensing \cite{WLFY12}, particle acceleration \cite{LCDJ11}, black hole complexity \cite{JDLi19}, and holography \cite{KLPe20}. The Taub-NUT solutions were proposed as black hole candidates with Misner string singularities on the axes. However, the physical interpretation of the NUT parameter still remains controversial \cite{taub51,NeTU63,Misn63,Hawk73,Haji71,MaRu05}.
Fortunately, recent researches show that the Misner string singularities are much less defective than {that} previously expected and the Taub-NUT solutions may actually be physically relevant \cite{CGGu15,CGGu16,CGGu18}.
On the other hand, black hole thermodynamics plays an important role in the study of the weak cosmic censorship conjecture. Even though black hole solutions with a NUT parameter have been found out for more than half a century, there {are} no convincing results for their thermodynamics \cite{Hawk1998a,Hawk98b,CEJM99,EmJM99,Mann99,Mann00,John14a,John14b,GaMa00,WuWu19}. Recently, the thermodynamics of black holes with a NUT parameter {seems} to have been reasonably formulated with the existence of Misner strings \cite{KuMa19,BoGK19,ChJi19}.

In this paper, we try to destroy the event horizon of the Reissner-Nordstr\"om Taub-NUT regular black hole by adopting the latest results of the black hole thermodynamics.  This black hole has no spacetime singularity, so it is not protected by the weak cosmic censorship conjecture. The destruction of its event horizon might provide us the possibility to access {regions inside the black hole} and give us useful information to build a consistent theory of quantum gravity.

The rest of this paper is organized as follows. In Sec.\,\ref{NUTbh}, we review the charged Taub-NUT black hole and its thermodynamics. In Sec.\,\ref{Field}, the scattering of a charged scalar field is explored in the charged Taub-NUT black hole background. In Sec.\,\ref{Thermo}, we study the conserved charges for the charged scalar field in the scattering process. We try to destroy the event horizon of the charged Taub-NUT black hole with a test scalar field  and a test particle in {Sec.\,\ref{DTEHBT} and Sec.\,\ref{WCCCP}}, respectively. The last section summarizes our results.

\section{the thermodynamics of the charged Taub-NUT Black hole }\label{NUTbh}

The charged Taub-NUT black hole is a solution of the Einstein-Maxwell theory. The metric of the charged Taub-NUT spacetime reads
\begin{align}
d s^{2}=&-\frac{f(r)}{r^{2}+n^{2}}(dt+2n\cos\theta d\phi)^{2} +\frac{r^{2}+n^{2}}{f(r)}dr^{2}
            + (r^{2}+n^{2})(d\theta^{2}+\sin^{2}\theta d\phi^{2}),\label{the NUT matric}
\end{align}
where
\begin{equation}
	f(r)=r^{2}-2Mr-n^{2}+e^{2} ,\label{f(r)}
\end{equation}
and the electromagnetic potential is
\begin{equation}\label{the electromagnetic gauge potential}
	\mathbf{A} =\frac{-e r}{r^{2}+n^{2}}({ d}t+2n\cos\theta  d\phi),
\end{equation}
where $M$, $n$, and $e$ denote the mass, the NUT parameter, and the electric parameter of the black hole, respectively. For the charged Taub-NUT black hole, there exist Misner string singularities located at $\theta=0$ and $\theta=\pi$, which can be seen from Fig. \ref{figure 1}. The Reissner-Nordstr\"om solution will be recovered from Eq. \eqref{the NUT matric} with the NUT parameter $n$ vanishing.
{For the charged Taub-NUT black hole, the scalar curvature  vanishes. To check if the black hole is regular, we obtain the {expression} of the Kretschmann scalar of this black hole as
\begin{equation}
\begin{aligned}
K=R^{\mu\nu\rho\tau}R_{\mu\nu\rho\tau}
&~~=\frac{8}{\left(n^2+r^2\right)^6}\left[e^4\left(7n^4-34n^2r^2+7r^4\right)\right.\\
&~-12e^2\left(n^6-10n^5r^2+5n^2r^4\right)-12e^2Mr\left(5n^4-10n^2r^2+r^4\right)\\
&~~+6\left(n^2-M^2\right)\left(n^2-r^2\right)\left(n^2-4nr+r^2\right)
\left(n^2+r^2\right)\\
&~~\left.\left(n^2+4nr+r^2\right)+24Mn^2r\left(n^2-3r^2\right)\left(3n^2-r^2\right)\right].
\end{aligned}
\end{equation}
The plot of the Kretschmann scalar is shown in Fig. \ref{1}. It shows that the Kretschmann scalar is regular everywhere.} So the charged Taub-NUT solution \eqref{the NUT matric} has no spacetime singularity {and} this black hole is regular and can be considered as a regular solution of the Reissner-Nordstr\"om black hole. {Furthermore, this black hole is geodesically complete and according to the classification in \cite{CDLV19}, it can be regarded as a one-way hidden wormhole. Therefore, after the destruction of the event horizon, there is no naked singularity and it is possible for us to explore regions inside the black hole and find new physical phenomenon. }

The horizons of the charged Taub-NUT black hole are determined by the equation
\begin{equation}\label{metric function}
	f(r)=r^{2}-2Mr-n^{2}+e^{2}=0.
\end{equation}
The outer and inner horizons can be obtained easily as
\begin{equation}\label{outer and inner horizons}
	r_{\pm}=M\pm \sqrt{M^{2}+n^{2}-e^{2}},
\end{equation}
where the $``+"$ sign corresponds the event horizon of the black hole and we denote it by $r_{h}$ in the following sections. When $M^{2}+n^{2}=e^{2}$, the two horizons coincide with each other and the black hole becomes an extremal one. The horizons of the black hole disappear for $M^{2}+n^{2}<e^{2}$.

\begin{figure*}
\subfigure{
\begin{minipage}[t]{0.45\linewidth}
\centering
\setcounter{figure}{0}
\includegraphics[width=3in]{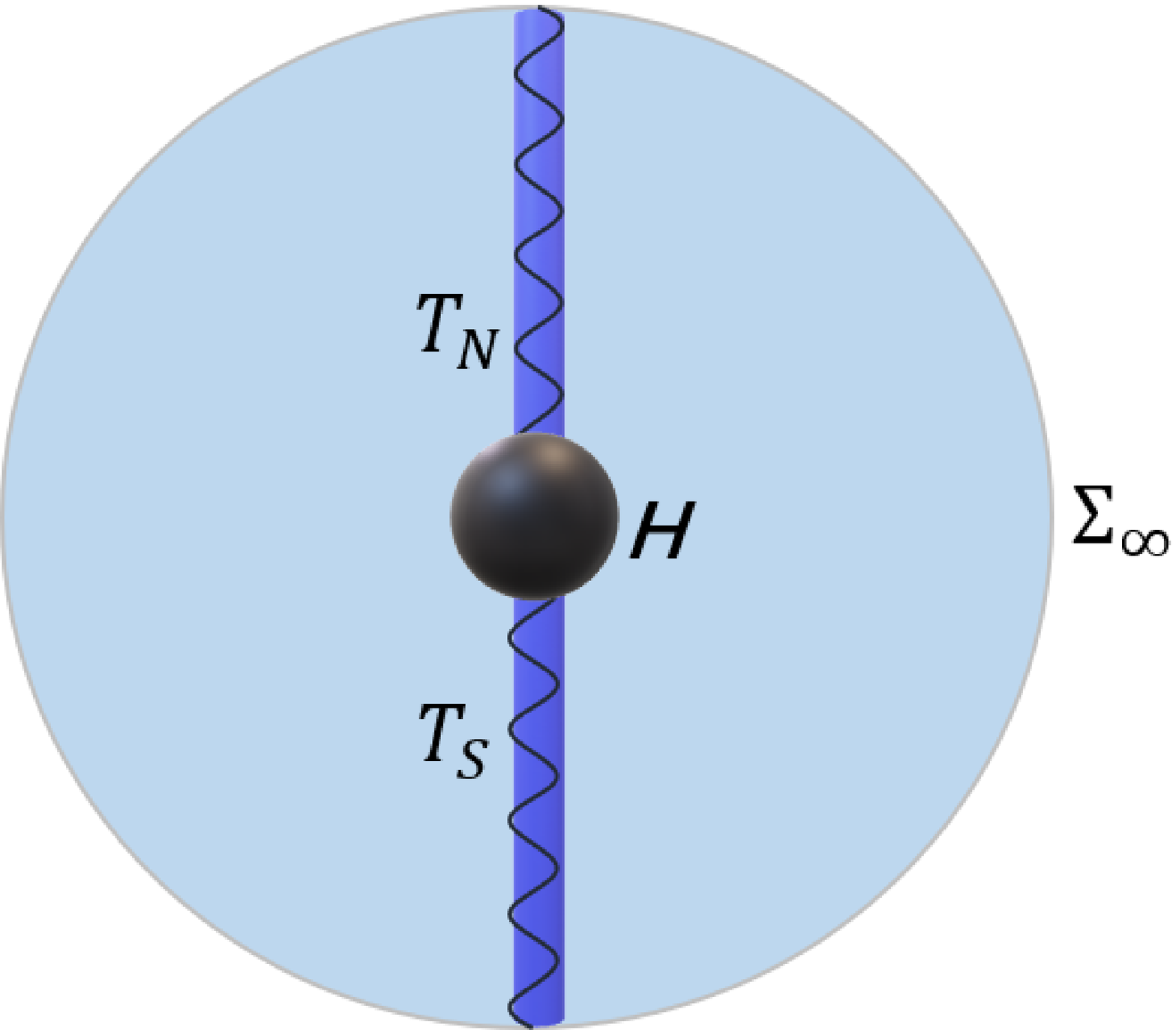}
\caption{Charged Taub-NUT boundaries \cite{BGHK19}: Misner tubes. The charged Taub-NUT space-time not only has the standard boundaries: the event horizon $H$ and spatial infinity $\Sigma_{\infty}$, but also includes two Misner tubes $T_{S}$ and $T_{N}$ located at the south and north axes, respectively.}
\end{minipage}
}\label{figure 1}
 \subfigure{
\begin{minipage}[t]{0.45\linewidth}
\centering
\includegraphics[width=3in]{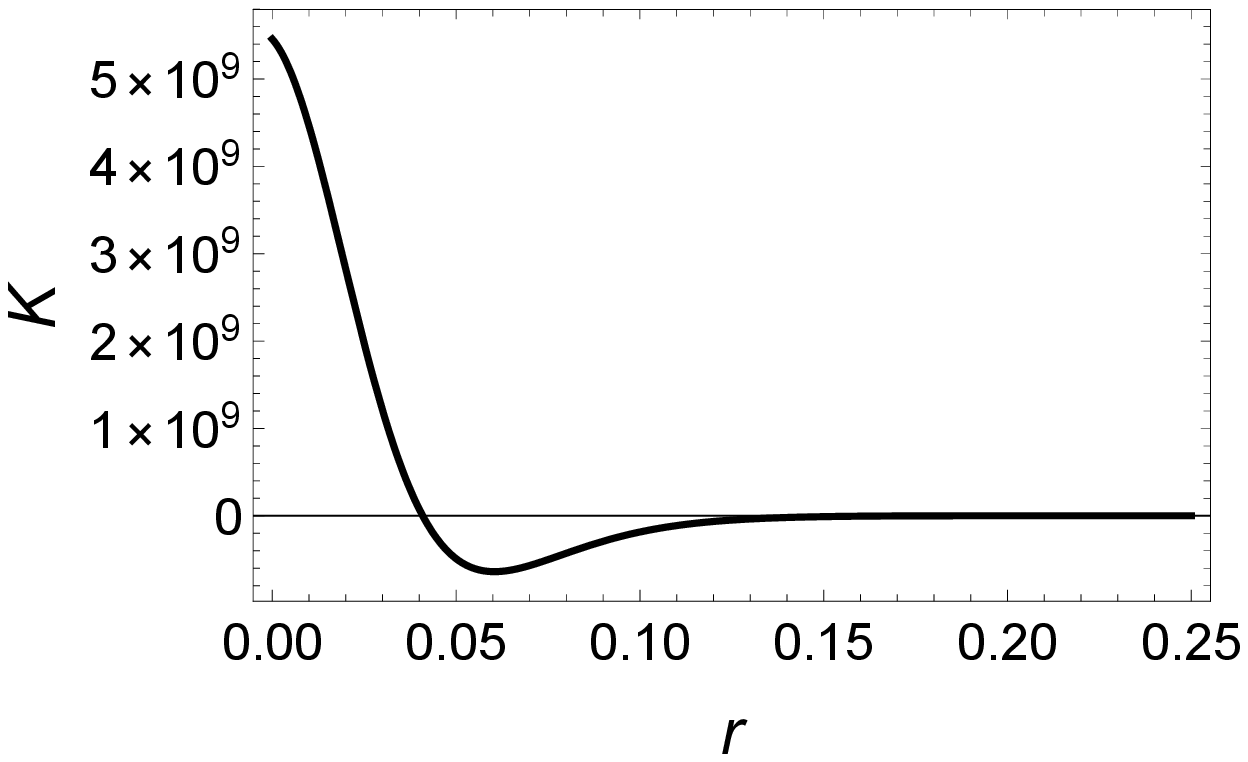}
\caption{{Plot of the Kretschmann scalar $K$ for a charged NUT black hole with the mass $M=1$, the charge parameter $e=1$, and the NUT parameter $n=0.1$.}}
\end{minipage}
}\label{1}
\end{figure*}

The Hawking temperature $T$ and the entropy $S$ of the black hole are
\begin{equation}
	T=\frac{1}{4\pi r_{\rm h}}\left(1-\frac{e^{2}}{r_{{\rm h}}^{2}+n^{2}}\right),\label{the Hawking temperature}
\end{equation}
and
\begin{equation}
	S=\pi(r_{\rm h}^{2}+n^{2}) .\label{the entropy}
\end{equation}
The electric potential is
\begin{equation}\label{electric charge and potential}
\Phi_{\rm h}=\frac{er_{\rm h}}{r_{\rm h}^{2}+n^{2}}.
\end{equation}

The first law of thermodynamics for the charged Taub-NUT black hole has been investigated extensively, and its expression is given by \cite{BoGK19}
\begin{equation}\label{The first law of thermodynamics}
	dM=TdS+\Phi_{\rm h} dQ+\Psi dN,
\end{equation}
where $Q$ is the electric charge surrounded by the event horizon
\begin{equation}
Q=\frac{e(r_{\rm h}^{2}-n^{2})}{r_{\rm h}^{2}+n^{2}},
\end{equation}
and the Misner charge $N$ is associated with the NUT parameter
\begin{equation}\label{the Misner charge}
N=-\frac{4\pi n^{3}}{r_{\rm h}}\left[1-\frac{e^{2}(n^{2}+3r_{\rm h}^{2})}{(n^{2}+r_{\rm h}^{2})^{2}}\right],
\end{equation}
with its conjugate quantity  $\Psi$
\begin{equation}\label{the Misner potential}
\Psi=\frac{1}{8\pi n}.
\end{equation}
{Here we want to point out that the conjugate quantity  $\Psi$ diverges as the NUT parameter $n\to 0$. This peculiar feature is similar to the thermodynamics of accelerated black holes \cite{ApGK17}.}

\section{Massive complex scalar field in charged Taub-NUT space-time}\label{Field}

In this section, we consider the scattering of a massive complex scalar field $\varphi$ minimally coupled to gravity in the charged Taub-NUT spacetime. The dynamics of the complex scalar field satisfies the Klein-Gordon equation
\begin{equation}
\frac{1}{\sqrt{-g}} (\partial_{\mu} - iqA_{\mu})\left[\sqrt{-g} g^{\mu \nu} (\partial_{\nu}-iqA_{\nu}) \varphi\right]-\mu_{s}^{2} \varphi=0,\label{the Klein-Gordon equation}
\end{equation}
where $\mu_{s}$ is the mass of the scalar field $\varphi$. Without loss of generality, the complex scalar field can be decomposed as \cite{RMCh20}
\begin{equation}
\varphi(t, r, \theta, \phi)=e^{-i \omega t} R(r) \Theta(\theta) e^{i m \phi},\label{the decomposition}
\end{equation}
where $\Theta(\theta)$ denotes a generalized spheroidal function. Inserting the metric \eqref{the NUT matric} and the decomposition \eqref{the decomposition} into the equation of motion \eqref{the Klein-Gordon equation}, we get the radial part of the equation
\begin{equation}
\frac{d}{d r}\left(f(r) \frac{d R(r)}{d r}\right)+\left(\frac{L^{2}}{f(r)}-\mu_{s}^{2} (r^{2}+n^{2})-\rho \right) R(r)=0,\label{the radial part}
\end{equation}
and the angular part \cite{BCCa06}
\begin{equation}
\frac{1}{\text{sin}\, \theta}\frac{d}{ d \theta}\left( \operatorname{sin} \theta \frac{d \Theta}{d \theta}\right)-\left[ \left(2\omega n \operatorname{cot}\theta -\frac{m}{\operatorname{sin}\theta} \right)^{2} -\rho\right] \Theta=0,\label{the angular part}
\end{equation}
where $\rho = l (l+1) + \mathcal{O}(l^{2})$ is the separation constant and
\begin{equation}
L= e\,q\,r +\omega (n^{2}+r^{2}). \label{G}
\end{equation}
Using the normalization condition \cite{Gwak18}
, the angular part $\Theta(\theta)$ can be reduced to unity in the charge and energy fluxes, so the exact expression of $\Theta(\theta)$ is inessential in this paper. Thus we will focus on the radial part function $R(r)$. To simplify the radial part equation \eqref{the radial part}, we introduce the tortoise coordinate
\begin{equation}
\frac{dr}{dr_{\ast}}=\frac{f(r)}{n^{2}+r^{2}}. \label{tortoise coordinate}
\end{equation}
Then Eq. \eqref{the radial part} becomes
\begin{align}\label{the radial part in tortoise coordinate}
\frac{d^{2} R(r)}{d r_{\ast}^{2}} +\frac{2r f(r)}{(r^{2}+n^{2})^{2}}\frac{d R(r)}{d r_{\ast}}  + \left[\left(\omega-\frac{e q r}{n^{2}+r^{2}}\right)^{2}-\frac{(\mu_{s}^{2}n^{2}+\mu_{s}^{2}r^{2}+\rho)f(r)}{(r^{2}+n^{2})^{2}} \right] R(r)=0.
\end{align}
Near the event horizon, $f(r) \to 0$ and Eq.~\eqref{the radial part in tortoise coordinate} can be reduced to
\begin{equation}\label{the radial part near horizon}
\frac{d^{2} R(r)}{d r_{\ast}^{2}} + \left( \omega - q \Phi_{\rm h}\right)^{2}  R(r)=0.
\end{equation}
The solution of the radial part can be solved as
\begin{equation}\label{solution of radial part}
   R(r) = e ^{\pm (\omega -q \Phi_{\rm h})r_{\ast}} .
\end{equation}
There are two branches of the solution. First branch with positive sign corresponds to the outgoing wave, and the second one with minus sign corresponds to the ingoing wave which is chosen as physically acceptable solution in this paper. Therefore, the wave function near the horizon is
\begin{equation}\label{wave function}
\varphi(t, r, \theta, \phi)=e^{-i \omega t} e ^{- (\omega -q \Phi_{\rm h})r_{\ast}} \Theta(\theta) e^{i m \phi}.
\end{equation}
With this function, we can study the changes of the black hole parameters after the ingoing wave scattering at the event horizon.

\section{Conserved charges under scattering of the scalar field}\label{Thermo}

In this paper, we neglect the self-force effect and other interactions, which means that the energy and electric charge carried by the wave must be small enough. The energy change of the black hole is related to the energy flux, which is determined by the energy-momentum tensor of the massive scalar field
\begin{equation}\label{energy-momentum tensor of scalar wave}
T^{\mu}_{\nu}=\frac{1}{2}D^{\mu}\varphi\partial_{\nu}\varphi^{\ast}+\frac{1}{2}D^{\ast\mu}\varphi^{\ast}\partial_{\nu}\varphi- \delta^{\mu}_{\nu}\left(\frac{1}{2}g^{\alpha\beta}D_{\alpha}\varphi D_{\beta}^{\ast}\varphi^{\ast} + \mu_{s}^{2}\varphi^{\ast}\varphi \right).
\end{equation}
With the energy-momentum tensor \eqref{energy-momentum tensor of scalar wave} and the ingoing wave function \eqref{wave function}, the energy flux through the event horizon is obtained by
\begin{equation}\label{energy flux}
\frac{d E}{d t}=\int_{H} T_{t}^{r} \sqrt{-g} d \theta d \phi=\omega\left(\omega-q \Phi_{\rm h}\right)\left(r_{\mathrm{h}}^{2}+n^{2}\right),
\end{equation}
and the charge flux is
\begin{equation}\label{charge flux}
\frac{d Q}{d t}=-\int_{H} j^{r} \sqrt{-g} d \theta d \phi=q \left(\omega-q \Phi_{\rm h}\right)\left(r_{\mathrm{h}}^{2}+n^{2}\right),
\end{equation}
where
the electric current $j^{\mu}$ is obtained by
\begin{equation}\label{the electric current}
j^{\mu}=-\frac{1}{2}i q [\varphi^{\ast}(\partial^{\mu}+i q A^{\mu})\varphi-\varphi(\partial^{\mu}-i q A^{\mu})\varphi^{\ast}].
\end{equation}
Due to the presence of the Misner strings, the integration turns to be more subtle. As shown in Fig. \ref{figure 1}, the event horizon $H$ does not contain the string singularities, which are located at $\theta=0$ and $\theta=\pi$, respectively. In Eqs. \eqref{energy flux} and \eqref{charge flux}, the normalization condition
\begin{align}\label{the normalization condition}
\lim\limits_{\varepsilon\to 0} \int^{\pi-\varepsilon}_{\varepsilon} \Theta^{2}(\theta) \operatorname{sin}\theta d \theta \int^{2\pi}_{0} d\phi
=\int^{\pi}_{0} \Theta^{2}(\theta) \operatorname{sin}\theta d \theta \int^{2\pi}_{0} d\phi =1
\end{align}
has been used. Here we have set the radii of both Misner tubes to be the same and equal $\varepsilon$ ($\varepsilon\ll 1$).

With the energy flux \eqref{energy flux} and the charge flux \eqref{charge flux}, we can get the changed energy and charge within a given infinitesimal time interval $dt$ as
\begin{equation}\label{transferred energy}
	d M =d E=\omega\left(\omega-q \Phi_{\rm h}\right)\left(r_{\mathrm{h}}^{2}+n^{2}\right) dt,
\end{equation}
\begin{equation}\label{transferred charge}
	d Q =q \left(\omega-q \Phi_{\rm h}\right)\left(r_{\mathrm{h}}^{2}+n^{2}\right) dt.
\end{equation}

From the above equations, we can see that the relation between $ \omega $ and $q \Phi_{\rm h}$ determines the direction of the energy flux and the charge flux. When $\omega > q \Phi_{\rm h}$, the energy and charge of the black hole increase. They remain unchanged when $\omega = q \Phi_{\rm h}$ while decrease when $\omega < q \Phi_{\rm h}$. The latter means that the energy and charge are extracted out by the scattering field, which is so called superradiance \cite{BCPa15}.

There are three parameters in the charged Taub-NUT black hole: the mass $M$, the electric parameter $e$, and the NUT parameter $n$. For a black hole far from extremal, its final state is still a black hole after exchanging the energy and charge. So we can use laws of black hole thermodynamics to investigate the changes of the three parameters during the scattering process as the authors did for Kerr-Taub-NUT black hole \cite{YCWWL20}.

If we assume that the test field only changes the energy and charge of the black hole, and the NUT parameter $n$ stays fixed in this process, through the first law of thermodynamics \eqref{The first law of thermodynamics}, we get
\begin{align}\label{The first law of thermodynamics if n fixed}
dS =\frac{1}{T}(dM-\Phi_{\rm h}dQ-\Psi dN) \nonumber =\frac{q^{2}r_{\rm h}^{2}}{T}\left(\frac{\omega}{q}-\frac{e}{r_{\rm h}}\right)
           \left(\frac{\omega}{q}-\frac{er_{\rm h}}{r_{\rm h}^{2}+n^{2}}\right)dt.
\end{align}
Since the NUT parameter $n$ is non-vanishing, it is clear that
\begin{equation}\label{charge if n fixed}
\frac{e}{r_{\rm h}}>\frac{er_{\rm h}}{r_{\rm h}^{2}+n^{2}}.
\end{equation}
For the wave modes with ${\omega}/{q}$ satisfying
\begin{equation}\label{wave modes if n fixed}
\frac{e}{r_{\rm h}}>\frac{\omega}{q}>\frac{er_{\rm h}}{r_{\rm h}^{2}+n^{2}},
\end{equation}
the entropy $S$ of the black hole decreases in the scattering process and this goes against the second law of thermodynamics \cite{Hawk72,CCFKNNB18}. Therefore, the NUT parameter $n$ should change in the scattering process.

Now we consider the case that the Misner charge $N$ is unchanged in the scattering process.
The change of the entropy is
\begin{equation}\label{The first law of thermodynamics if N fixed}
dS =\frac{1}{T}(dM-\Phi_{\rm h}dQ-\Psi dN)= \frac{1}{T}(\omega-q\Phi_{\rm h})^{2}dt\ge 0.
\end{equation}
The result shows that the entropy never decreases and this satisfies the second law of thermodynamics of the black hole. So we will consider that the Misner charge $N$ is conserved in the following discussion.

\section{destroying the event horizon with test scalar field}\label{DTEHBT}

In this section, we try to destroy the black hole event horizon by scattering a classical complex scalar field into an extremal and a near-extremal charged Taub-NUT black hole, respectively. The energy and charge of the black hole will change after scattering the test scalar field. If the black hole is overcharged, its event horizon will disappear and the inner structure of the black hole might be seen.

To guarantee the existence of the event horizon, the minimum value of the metric function $f  (r)$ must be non-positive. By using this condition, we can check whether the black hole event horizon is destroyed. For an extremal or {a} near-extremal charged Taub-NUT black hole, the minimum value of the metric function $f(r)$ is
\begin{equation}\label{minimum of near-extremal}
	f_{\rm min}=-M^{2}+e^{2}-n^{2},
\end{equation}
and the corresponding coordinate $r$ is located at $r_{0}=M$. The initial state of the black hole at the minimum value is represented by $f_{\rm min}(M, Q, N)$. After the scattering of the scalar field, the parameters of the final state become
\begin{equation}\label{the parameters change}
	\begin{aligned}
		M & \rightarrow M^{\prime}=M+d M, \\
		Q & \rightarrow Q^{\prime}=Q+d Q, \\
		N & \rightarrow N^{\prime}=N,
	\end{aligned}
\end{equation}
where we have supposed that the Misner charge $N$ is unchanged during the scattering process. The minimum value of the final state metric function $f_{\rm min}(M+d M, Q+d Q, N)$ can be expressed in term of the initial state function {$f_{\rm min}(M,Q, N)$},
\begin{align}
	f_{\rm min}^{\rm{final}} &\equiv
      f_{\rm min}(M+d M, Q+d Q, N)  \nonumber \\
      &= f_{\rm min}(M,Q, N) + \left(\frac{\partial f_{\rm min} }{\partial M} \right)_{Q,N}dM + \left(\frac{\partial f_{\rm min} }{\partial Q} \right)_{M,N}dQ , \label{express of final state}
\end{align}
where
\begin{equation}\label{derivatives}
\begin{aligned}
	\left(\frac{\partial f_{\rm min} }{\partial M} \right)_{Q,N}&= \frac{2[n^{4}-n^{2}r_{\rm h}^{2}-M^{2}(n^{2}-9r_{\rm h}^{2})+6Mr_{\rm h}(n^{2}-r_{\rm h}^{2})]}{-6n^{2}r_{\rm h}+6r_{\rm h}^{3}+M(n^{2}-9r^{2}_{\rm h})},\\
	\left(\frac{\partial f_{\rm min} }{\partial Q} \right)_{M,N}&=\frac{2e[2r^{3}_{\rm h}(n^{2}-3r_{\rm h}^{2})+M(n^{4}+2n^{2}r^{2}_{\rm h}+9r_{\rm h}^{4})]}{(n^{2}+r_{\rm h}^{2})(-Mn^{2}+6n^{2}r_{\rm h}+9Mr^{2}_{\rm h}-6r^{3}_{\rm h})}.
\end{aligned}
\end{equation}
When the minimum value of the final state metric function $f_{\rm min}^{\rm{final}}$ becomes positive, the black hole will be overcharged in the scattering process and its event horizon will be destroyed.

For an extremal charged Taub-NUT black hole, the metric function $f(r)$ has only one intersection with the $r$-axis and this means that the localization of the event horizon coincides with the one of the minimum value, i.e. $r_{\rm h}=r_{0}=M$. Then the minimum value of the final state metric function $f_{\rm min}^{\rm{final}}$ \eqref{express of final state} can be obtained as
\begin{equation}\label{minimum value for extramal black hole}
\begin{aligned}
f_{\rm min}^{\rm{final}}=-\frac{2(n^{4}+4n^{2}r^{2}_{\rm h}+3r_{\rm h}^{4})}{5r_{\rm h}n^{2}+3r_{\rm h}^{3}}(\omega-q\Phi_{\rm h})^{2}(n^{2}+r_{\rm h}^{2})dt\le 0.
\end{aligned}
\end{equation}
Here the equal sign is taken only when $\omega=q\Phi_{\rm h}$ and this means that the final state is still an extremal black hole. For $\omega\neq q\Phi_{\rm h}$, the result shows that the extremal black hole will become a non-extremal one after absorbing the test scalar field. Thus, the event horizon of the extremal charged Taub-NUT black hole cannot be destroyed after the scattering of the test scalar field.

Now we consider the near-extremal charged Taub-NUT black hole case by using the above method.
With the transferred energy \eqref{transferred energy} and the transferred charge \eqref{transferred charge} during the time interval $dt$, the minimum value of the final state metric function $f_{\rm min}^{\rm{final}}$ \eqref{express of final state} is given by
\begin{equation}\label{minimum value for the metric function}
f_{\rm min}^{\rm{final}}=-M^{2}+e^{2}-n^{2}+\Gamma(\omega-q\tilde{\Phi})(\omega-q\Phi_{\rm h})(n^{2}+r^{2}_{\rm h})dt,
\end{equation}
where
\begin{equation}\label{Gamma}
\Gamma=\frac{2[n^{4}-n^{2}r^{2}_{\rm h}-M^{2}(n^{2}-9r_{\rm h}^{2})+6Mr_{\rm h}(n^{2}-r^{2}_{\rm h})]}{-6n^{2}r_{\rm h}+6r_{\rm h}^{3}+M(n^{2}-9r^{2}_{\rm h})},
\end{equation}
and
\begin{equation}\label{Phi prime}
\tilde{\Phi}=\Phi_{\rm h}\frac{2r_{\rm h}^{3}(n^{2}-3r^{2}_{\rm h})+M(n^{4}+2n^{2}r_{\rm h}^{2}+9r_{\rm h}^{4})}{n^{4}r_{\rm h}-n^{2}r_{\rm h}^{3}-M^{2}r_{\rm h}(n^{2}-9r^{2}_{\rm h})+6Mr^{2}_{\rm h}(n^{2}-r_{\rm h}^{2})}.
\end{equation}

Since the distance between the minimum point $r_{0}$ and the event horizon radius $r_{\rm h}$ can be extremely small for a near-extremal black hole, we can set $r_{\rm h}=r_{\rm 0}+\epsilon$, where $0< \epsilon \ll 1$. We also set $dt \sim \epsilon$ because the process considered here occurs in an infinitesimal time interval. Then Eq. \eqref{minimum value for the metric function} can be rewritten as
\begin{equation}\label{minimum value for near extramal black hole}
f_{\rm min}^{\rm{final}}=-\epsilon^{2}+\Gamma q^{2}\left(\frac{\omega}{q}-\tilde{\Phi}\right)\left(\frac{\omega}{q}-\Phi_{\rm h}\right)(n^{2}+r^{2}_{\rm h})\epsilon.
\end{equation}
Equation \eqref{minimum value for near extramal black hole} can be regarded as a quadratic function in term of ${\omega}/{q}$.
Due to
\begin{eqnarray}\label{Gamma is small that 0}
\Gamma &=& -\frac{2[n^{4}+n^{2}(4r^{2}_{\rm h}-4r_{\rm h}\epsilon-\epsilon^{2})+3r^{2}_{\rm h}(r^{2}_{\rm h}-4r_{\rm h}\epsilon+3\epsilon^{2})]}{3r_{\rm h}^{2}(r_{\rm h}-3\epsilon)+n^{2}(5r_{\rm h}+\epsilon)} \nonumber \\
 &=&-\frac{2(n^{4}+ 4n^{2}r^{2}_{\rm h}+3r^{4}_{\rm h})}
          {3r_{\rm h}^{3} + 5 n^{2}r_{\rm h}}
    + \mathcal{O}(\epsilon)
<0,
\end{eqnarray}
if the maximum value of the final state $f_{\rm min}^{\rm{final}}$ is nonpositive, the other values of the final state are always negative, which means that the event horizon of the black hole still exists. When the wave mode with ${\omega}/{q}$ satisfies
\begin{equation}\label{wave mode}
\frac{\omega}{q}=\frac{\tilde{\Phi}+\Phi_{\rm h}}{2},
\end{equation}
$f_{\rm min}^{\rm{final}}$ is a maximum and can be expressed as
\begin{equation}\label{the maximum value}
f_{\rm min}^{\rm{final}} = -\epsilon^{2}
   +\frac{q^{2}(n^{2}-3r^{2}_{\rm h})^{2}(n^{2}+r^{2}_{\rm h})^{2}\Phi_{\rm h}^{2}}
         {2r^{3}_{\rm h}(5n^{4}+18n^{2}r_{\rm h}^{2}+9r^{4}_{\rm h})}
    \epsilon^{3}
   <0,
\end{equation}
where the higher-order terms of $\epsilon$ have been omitted. Therefore, it is clear that the minimum value of the metric function $f_{\rm min}^{\rm{final}}$ \eqref{minimum value for near extramal black hole} is always negative. Furthermore, we can find that $0>f_{\rm min}^{\rm{final}}>f_{\rm min}$, which means that the final state is still a near-extremal black hole and it is closer to the extremal state than the initial state. The result shows that the event horizon still exists and the black hole cannot be overcharged. So the event horizon of the near-extremal NUT black hole cannot be destroyed after the scalar field scattering and the final state of this black hole is still a black hole.

Thus, both near-extremal and extremal Reissner-Nordstr$\ddot{\text{o}}$m Taub-NUT black holes cannot be overcharged and their horizons will not disappear after the scattering of the test scalar field.

\section{destroy the event horizon with test particle}\label{WCCCP}

Another method to destroy the black hole event horizon is throwing a test charged particle into a near-extremal or extremal charged Taub-NUT black hole. This gedanken experiment was proposed by Wald for the first time \cite{Wald74}. By ignoring back reaction, Hubeny found that the near-extremal Reissner-Nordstr$\ddot{\text{o}}$m black hole can be overcharged after capturing the test particle \cite{Hube99}. Further, Jacobson and Sotiriou found that the event horizon of the near-extremal Kerr black hole can be destroyed by the test particle without considering radiative and self-force effects \cite{JaSo09}. In this section, we will examine whether the event horizon of the charged Taub-NUT black hole still exists after absorbing a test charged particle.

The Lagrangian of a test particle with rest mass $\mu_{m}$ and charge $\delta Q$ is
\begin{equation}\label{The Lagrangian of a test particle}
L=\frac{1}{2} \mu_{m} g_{\mu \nu} \frac{d x^{\mu}}{d \tau} \frac{d x^{\nu}}{d \tau}+\delta Q A_{\mu}\frac{dx^{\mu}}{d\tau},
\end{equation}
where $\tau$ is the proper time of the test particle. From the Lagrangian \eqref{The Lagrangian of a test particle}, the equation of motion of the test particle can be derived as
\begin{equation}\label{EOM of a test particle}
\frac{d^{2} x^{\mu}}{d \tau^{2}}+\Gamma_{\alpha \beta}^{\mu} \frac{d x^{\alpha}}{d \tau} \frac{d x^{\beta}}{d \tau}=\frac{\delta Q}{\mu_{m}}F^{\mu}_{~\nu}\frac{dx^{\nu}}{d\tau}.
\end{equation}

In this paper, we consider that the particle is dropped on the equatorial plane ($\theta = \pi/2$) and along the radial direction, so the components $P_{\phi}$ and $P_{\theta}$ of the angular momentum of the particle vanish.
The energy and the angular momentum  are
\begin{align}
\delta E &=-\frac{\partial L}{\partial \dot{t}}=-\mu_{m} g_{t \nu} \frac{d x^{\nu}}{d \tau}-\delta QA_{t}, \label{energy change of a test particle}\\
P_{\phi} &=\frac{\partial L}{\partial \dot{\phi}}=\mu_{m} g_{\phi \nu} \frac{d x^{\nu}}{d \tau}+\delta QA_{\phi}=0, \label{angular momentum phi of a test particle}\\
P_{\theta}&= \frac{\partial L}{\partial \dot{\theta}}=\mu_{m} g_{\theta \theta} \frac{d \theta}{d \tau}=0. \label{angular momentum theta of a test particle}
\end{align}

Then we investigate the conditions to destroy the event horizon of the black hole. Obviously, the test particle should be able to enter into the event horizon and the black hole will be overcharged after absorbing the test particle, which can give the relations between the energy $\delta E$ and the charge $\delta Q$ of the test particle.

The condition for the test particle entering the event horizon requires that its motion outside the event horizon is timelike and future directed, i.e.
\begin{equation}\label{timelike condition}
g_{\mu\nu}\frac{dx^{\mu}}{d\tau}\frac{dx^{\nu}}{d\tau}=-1,
\end{equation}
\begin{equation}\label{future directed condition}
\frac{dt}{d\tau} >0.
\end{equation}

Substituting Eqs. \eqref{energy change of a test particle}, \eqref{angular momentum phi of a test particle}, and \eqref{angular momentum theta of a test particle} into Eq. \eqref{timelike condition}, we get
\begin{align}\label{the solution of particle energy}
&g_{\phi\phi}\delta E^{2}+2(A_{t}g_{\phi\phi}-A_{\phi}g_{t\phi})\delta Q\delta E +(A_{\phi}^{2} g_{tt}-2 A_{t}A_{\phi}g_{t\phi}+ A_{t}^{2}g_{\phi\phi}) \delta Q^{2}\nonumber \\
&=(\mu_{m}^{2}+g^{rr} P_{r}^{2})(g_{t\phi}^{2}-g_{tt}g_{\phi\phi})   .
\end{align}
The energy of the particle can be obtained as
\begin{align}\label{delta E}
\delta E&=\frac{1}{g_{\phi\phi}}\delta Q A_{\phi}g_{t\phi}-\delta Q A_{t} +
 \frac{1}{g_{\phi\phi}}
 \sqrt{(g_{t\phi}^{2}-g_{tt}g_{\phi\phi})
 	   (\delta Q^{2}A_{\phi}^{2}+\mu_{m}^{2}g_{\phi\phi}+g_{\phi\phi}g^{rr}P_{r}^{2})}.
\end{align}

Since the trajectory of the charged particle outside the event horizon should be future directed \eqref{future directed condition}, the condition of the test particle entering into the event horizon can be obtained as
\begin{equation}\label{the condition on the horizon for particle}
\delta E>-A_{t} \delta Q={\Phi_{\rm h}}\delta Q \equiv \delta E_{\rm min} .
\end{equation}

Furthermore, the condition for the black hole being overcharged requires that the minimum value of the metric function $f_{\rm min}^{\rm{final}}$ is positive. From Eq. \eqref{express of final state}, this condition to first order is
\begin{equation}\label{first order of overspin condition}
f_{\rm min}^{\rm{final}}
 = f_{\rm min}(M, Q, N) + \Gamma \delta M- \Gamma \tilde{\Phi}\delta Q >0.
\end{equation}
Noting that $\Gamma <0$ and $ f_{\rm min}\leq 0$, the condition \eqref{first order of overspin condition} can be rewritten as
\begin{equation}\label{rewritten first order of overspin condition}
\delta E< \tilde{\Phi} \delta Q-\frac{ f_{\rm min}}{\Gamma} \equiv \delta E_{\rm max}.
\end{equation}

When both conditions \eqref{the condition on the horizon for particle} and \eqref{rewritten first order of overspin condition} are satisfied simultaneously, the event horizon of the black hole will be destroyed.

For an extremal charged Taub-NUT black hole, we have $ f_{\rm min}=0$ and $r_{\rm h}=r_{0}=M$. With the expression of $\tilde{\Phi}$ in Eq.~\eqref{Phi prime}, we can get
\begin{equation}\label{Phi and Phi prime for extremal black hole}
\tilde{\Phi}=\Phi_{\rm h}.
\end{equation}
Therefore, we have
\begin{equation}\label{the conditions for extremal black hole}
\delta E_{\rm min}=\Phi_{\rm h}\delta Q=\delta E_{\rm max}.
\end{equation}
It is clear that both conditions \eqref{the condition on the horizon for particle} and \eqref{rewritten first order of overspin condition} cannot be satisfied simultaneously. This means that the test charged particle which is able to destroy the event horizon will feel repulsive force from the black hole and cannot enter into the event horizon. Thus, the event horizon of the extremal charged Taub-NUT black hole cannot be destroyed by the test charged particle.

For a near-extremal charged Taub-NUT black hole, the metric function $f_{\rm min}<0$. The conditions \eqref{the condition on the horizon for particle} and \eqref{rewritten first order of overspin condition} become
\begin{equation}\label{condition inequations 1}
\delta E >{ \Phi_{\rm h}}\delta Q=\delta E_{\rm min},
\end{equation}
\begin{equation}\label{condition inequations 2}
\delta E< \tilde{\Phi} \delta Q-\frac{ f_{\rm min}}{\Gamma} =\delta E_{\rm max},
\end{equation}
where $\tilde{\Phi}$ can be rewritten as
\begin{equation}\label{rewitten phi tilde}
\tilde{\Phi} = { \Phi_{\rm h}} +\frac{2 { r_{\rm h}} { \Phi_{\rm h}} +Q}{n^{2}+3{ r^2_{\rm h}}}\epsilon+\mathcal{O}(\epsilon^{2})>{ \Phi_{\rm h}}.
\end{equation}
There exists an available range of the energy $\delta E$ and the charge $\delta Q$ for the test particle satisfying the two inequations \eqref{condition inequations 1} and \eqref{condition inequations 2} to destroy the event horizon of the black hole. Plot of this range is shown in Fig. \ref{E and Q range}. As a result, the event horizon of the near-extremal charged Taub-NUT black hole can be destroyed for some charged particles.

The results show that the event horizon cannot be destroyed for an extremal charged Taub-NUT black hole, while can be destroyed for a near-extremal one.
We emphasize here that the weak cosmic censorship conjecture is not violated although the event horizon is destroyed. {The reason is that the charged Taub-NUT black hole is regular and there is no spacetime singularity in the whole spacetime, although the existence of the string singularities which are coordinate singularities rather than spacetime singularities}. Therefore, no naked singularity will appear after destroying the event horizon and the interior of the black hole might be accessible to distant observers.
\begin{figure*}
\centering
\setcounter{figure}{2}
\includegraphics[scale=1.0]{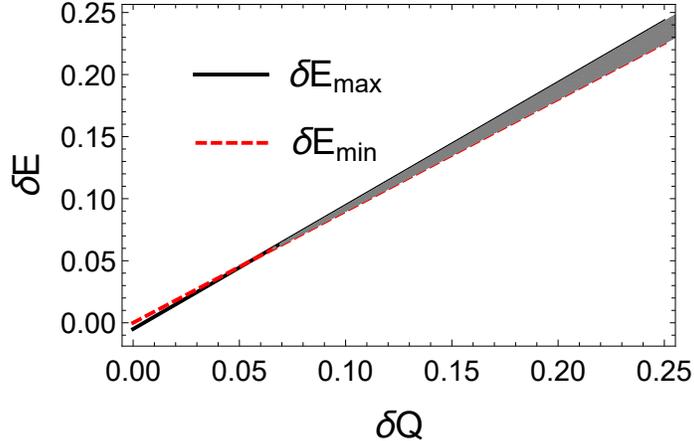}
	\caption{(color online) Energy bounds $\delta E_{\rm max}$ (black solid line) and $\delta E_{\rm min}$ (red dashed line) to destroy the event horizon of a near-extremal charged Taub-NUT black hole by dropping a particle with charge $\delta Q$. Here we have set the mass $M=1$, the charge parameter $e=1$, and the NUT parameter $n=0.1$. The grey region stands for $\delta E_{\rm max}>\delta E_{\rm min}$.
}
	\label{E and Q range}
\end{figure*}

\section{conclusion}\label{conclusion}

The weak cosmic censorship conjecture has become one of the foundations of black hole physics. It might ultimately turn out to be true. However, the destruction of the event horizon of a regular black hole does not lead to the appearance of naked singularity and this does not cause the loss of predictability. The disappearance of the event horizon of a regular black hole is not forbidden by the weak cosmic censorship conjecture, and this might provide us the possibility to access regions behind the event horizon and might provide us observable information to build a consistent theory of quantum gravity.

In this paper, we investigated the possibility of destroying the event horizon of the charged Taub-NUT black hole by a test charged scalar field and a test charged particle, respectively. For the test scalar field scattering, both the near-extremal and extremal charged Taub-NUT black holes cannot be overcharged. For the test particle absorption, the result suggests that the event horizon of the extremal charged Taub-NUT black hole cannot be destroyed while the near-extremal charged Taub-NUT black hole can be overcharged. However, if we assume that the test particle crosses the event horizon of the black hole in a continuous path, i.e. the charge and energy are transferred to the black hole gradually, the horizon may be still stable as indicated by Gwak \cite{Gwak17}.

\section*{Acknowledgement}

{We acknowledge Shao-Wen Wei for useful suggestions and thank Jun-Jie Wan for many inspiring discussions. {Particularly, we appreciate {the} referees' patience and comments and also thank Yu-Peng Zhang for his invaluable help.} This work was supported by the National Natural Science Foundation of China (Grants No. 11875151 and No. 11522541) and the Fundamental Research Funds for the Central Universities (Grants Nos. lzujbky2019-it21 and No. lzujbky-2019-ct06).}


\begin{thebibliography}{99}
{

\bibitem{LIGO16}
B.~P.~Abbott \textit{et al.} [LIGO Scientific and Virgo],
{ Observation of gravitational waves from a binary black hole merger},
Phys.\ Rev.\ Lett.\  {\bf 116}, 061102 (2016).

\bibitem{LIGO17}
B.~P.~Abbott \textit{et al.} [LIGO Scientific and Virgo],
{  GW170817: observation of gravitational waves from a binary neutron star inspiral},
Phys.\ Rev.\ Lett.\  {\bf 119}, 161101  (2017).

\bibitem{LIGO20}
B.~P.~Abbott \textit{et al.} [LIGO Scientific and Virgo],
{  GW190521: a binary black hole merger with a total mass of $150  M_{\odot}$},
Phys.\ Rev.\ Lett.\  {\bf 125}, 101102  (2020).

\bibitem{Zhan19}
X.~Zhang,
{  Gravitational wave standard sirens and cosmological parameter measurement},
Sci.\ China-Phys.\ Mech.\ Astron.\ {\bf 62}, 110431 (2019).

\bibitem{LCHu19}
J.~Li, Z.~C.~Che, and Q.~G.~Huang,
{  Measuring the tilt of primordial gravitational-wave power spectrum from observations},
Sci.\ China-Phys.\ Mech.\ Astron.\ {\bf 62}, 110421 (2019).

\bibitem{CFMPR17}
V.~Cardoso, E.~Franzin, A.~Maselli, P.~Pani and G.~Raposo,
{  Testing strong-field gravity with tidal Love numbers},
Phys.\ Rev.\ D {\bf 95}, 084014  (2017).

\bibitem{BDFM18}
E.~Belgacem, Y.~Dirian, S.~Foffa and M.~Maggiore,
{  Modified gravitational-wave propagation and standard sirens},
Phys.\ Rev.\ D {\bf 98}, 023510 (2018).

\bibitem{FLLZC19}
X.~L.~Fan, J.~Li, X.~Li, Y.~H.~Zhong, and J.~W.~Cao,
{  Applying deep neural networks to the detection and space parameter estimation of compact binary coalescence with a network of gravitational wave detectors},
Sci.\ China-Phys.\ Mech.\ Astron.\ {\bf 62}, 969512 (2019).

\bibitem{GBPU16}
J.~Garcia-Bellido, M.~Peloso and C.~Unal,
{ Gravitational waves at interferometer scales and primordial black holes in axion inflation},
JCAP {\bf 12}, 031 (2016).

\bibitem{NiZh19}
R.~Niu and W.~Zhao,
{ Constraining the non-Einsteinian polarizations of gravitational waves by pulsar timing array},
Sci.\ China-Phys.\ Mech.\ Astron.\ {\bf 62}, 970411 (2019).

\bibitem{Heis18}
L.~Heisenberg,
{ A systematic approach to generalisations of General Relativity and their cosmological implications},
Phys. Rept. {\bf 796}, 1 (2019).

\bibitem{Isha18}
M.~Ishak,
{ Testing General Relativity in Cosmology},
Living Rev. Rel.  {\bf 22}, 1 (2019).

\bibitem{FSCMMF17}
W.~M.~Farr, S.~Stevenson, M.~Coleman Miller, I.~Mandel, B.~Farr and A.~Vecchio,
{ Distinguishing Spin-Aligned and Isotropic Black Hole Populations With Gravitational Waves},
Nature  {\bf 548}, 426 (2017).

\bibitem{HJCa19}
X.~K.~He, J.~L.~Jing, and Z.~J.~Cao,
{  Generalized gravitomagnetic field and gravitational waves},
Sci.\ China-Phys.\ Mech.\ Astron.\ {\bf 62}, 110422 (2019).

\bibitem{IGFST19}
M.~Isi, M.~Giesler, W.~M.~Farr, M.~A.~Scheel and S.~A.~Teukolsky,
{ Testing the no-hair theorem with GW150914},
Phys.\ Rev.\ Lett.\  {\bf 123}, 111102  (2019).








\bibitem{LMLL20}
H.~S.~Liu, Z.~F.~Mai, Y.~Z.~Li, and H.~L$\ddot{\text{u}}$,
{ Quasi-topological electromagnetism: dark energy, dyonic black holes, stable photon spheres and hidden electromagnetic duality},
Sci.\ China-Phys.\ Mech.\ Astron.\ {\bf 63}, 240411 (2020).

\bibitem{KuMa19}
R.~A.~Hennigar, D.~Kubiz\v{n}\'{a}k, and R.~B.~Mann,
{ Thermodynamics of Lorentzian Taub-NUT spacetimes},
Phys.\ Rev.\ D\ \textbf{100}, 064055 (2019).

\bibitem{WuWu19}
S.~Q.~Wu and D.~Wu,
{ Thermodynamical hairs of the four-dimensional Taub-Newman-Unti-Tamburino spacetimes},
Phys.\ Rev.\ D\ \textbf{100}, 101501(R) (2019).


\bibitem{MiXv19}
Y.~G.~Miao and Z.~M.~Xu,
{ Interaction potential and thermo-correction to the equation of state for thermally stable Schwarzschild anti-de Sitter black holes},
Sci.\ China-Phys.\ Mech.\ Astron.\ {\bf 62}, 010412 (2019).

\bibitem{BoGK19}
A.~B.~Bordo, F.~Gray, and D.~Kubiz\v{n}\'{a}k,
{ Thermodynamics and phase transitions of NUTty dyons},
JHEP \textbf{07}, 119 (2019).


\bibitem{ChJi19}
Z.~Chen and J.~Jiang,
{ General Smarr relation and first law of a NUT dyonic black hole},
Phys.\ Rev.\ D\ \textbf{100}, 104016 (2019).

\bibitem{WWYu19}
Y.~Wang, C.~H.~Wu, and R.~H.~Yue,
{ A new rotating black hole in quintessential dark energy and its thermodynamics},
Sci.\ China-Phys.\ Mech.\ Astron.\ {\bf 62}, 110411 (2019).
}

\bibitem{Penr69}
R.~Penrose,
{ Gravitational collapse: the role of general relativity},
Riv.\ Nuovo Cim.\  {\bf 1}, 252 (1969).
[Gen.\ Rel.\ Grav.\  {\bf 34}, 1141 (2002)].




\bibitem{East19}
W.~E.~East,
{ Cosmic censorship upheld in spheroidal collapse of collisionless matter},
Phys.\ Rev.\ Lett.\  {\bf 122}, 231103 (2019).


\bibitem{FiKT16}
P.~Figueras, M.~Kunesch, and S.~Tunyasuvunakool,
{ End point of black ring instabilities and the weak cosmic censorship conjecture},
Phys.\ Rev.\ Lett.\  {\bf 116}, 071102 (2016).


\bibitem{FKLT17}
P.~Figueras, M.~Kunesch, L.~Lehner, and S.~Tunyasuvunakool,
{ End point of the ultraspinning instability and violation of cosmic censorship},
Phys.\ Rev.\ Lett.\  {\bf 118}, 151103 (2017).


\bibitem{CrSa17}
T.~Crisford and J.~E.~Santos,
{ Violating the weak cosmic censorship conjecture in four-dimensional anti-de Sitter space},
Phys.\ Rev.\ Lett.\  {\bf 118}, 181101 (2017).


\bibitem{HSSLW20}
T.~T.~Hu, Y.~Song, S.~Sun, H.~B.~Li, and Y.~Q.~Wang,
{ Weak cosmic censorship in Born-Infeld electrodynamics and bound on charge-to-mass ratio},
Eur. Phys. J. C \textbf{80}, 147 (2020).


\bibitem{SoHW20}
Y.~Song, T.~T.~Hu, and Y.~Q.~Wang,
{ Weak cosmic censorship with self-interacting scalar and bound on charge to mass ratio},
JHEP {\bf 045}, 03 (2021).




\bibitem{Wald74}
R.~M.~Wald,
{ Gedanken experiments to destroy a black hole},
Ann.\ Phys.\ \textbf{83}, 548 (1974).




\bibitem{RoCa11}
J.~V.~Rocha and V.~Cardoso,
{ Gravitational perturbation of the BTZ black hole induced by test particles and weak cosmic censorship in AdS spacetime},
Phys.\ Rev.\ D {\bf 83}, 104037 (2011).


\bibitem{LCNR10}
M.~Bouhmadi-Lopez, V.~Cardoso, A.~Nerozzi, and J.~V.~Rocha,
{ Black holes die hard: can one spin-up a black hole past extremality?}
Phys.\ Rev.\ D {\bf 81}, 084051 (2010).


\bibitem{ShDA19}
S.~Shaymatov, N.~Dadhich, and B.~Ahmedov,
{ The higher dimensional Myers-Perry black hole with single rotation always obeys the cosmic censorship conjecture},
Eur.\ Phys.\ J.\ C {\bf 79}, 585 (2019).


\bibitem{Hube99}
V.~E.~Hubeny,
{  Overcharging a black hole and cosmic censorship},
Phys.\ Rev.\ D {\bf 59}, 064013 (1999).


\bibitem{JaSo09}
T.~Jacobson and T.~P.~Sotiriou,
{ Over-spinning a black hole with a test body},
Phys.\ Rev.\ Lett.\  {\bf 103}, 141101 (2009);
Erratum: [Phys.\ Rev.\ Lett.\  {\bf 103}, 209903 (2009)].


\bibitem{LiBa13}
Z.~Li and C.~Bambi,
{ Destroying the event horizon of regular black holes},
Phys. Rev. D \textbf{87}, 124022 (2013).

\bibitem{SoWa17}
J.~Sorce and R.~M.~Wald,
{ Gedanken experiments to destroy a black hole. II. Kerr-Newman black holes cannot be overcharged or overspun},
Phys.\ Rev.\ D \textbf{96}, 104014 (2017).

\bibitem{JiGa20}
J.~Jiang and Y.~Gao,
{ Investigating the gedanken experiment to destroy the event horizon of a regular black hole},
Phys. Rev. D \textbf{101}, 084005 (2020).

\bibitem{QYWR20}
F.~Qu, S.~J.~Yang, Z.~Wang, and J.~R.~Ren,
{ Weak cosmic censorship conjecture is not violated for a rotating linear dilaton black hole},
[arXiv:2008.09950 [gr-qc]].



\bibitem{ZhJi20}
M.~Zhang and J.~Jiang,
{ New gedanken experiment on higher-dimensional asymptotically AdS Reissner-Nordstr\"om black hole},
Eur. Phys. J. C \textbf{80},  890 (2020).



\bibitem{Jian20}
J.~Jiang,
{ Static charged Gauss-Bonnet black holes cannot be overcharged by the new version of gedanken experiments},
Phys.\ Lett.\ B\ \textbf{804}, 135365 (2020).

\bibitem{JiZh20}
J.~Jiang and M.~Zhang,
{ New version of the gedanken experiments to test the weak cosmic censorship in charged dilaton-Lifshitz black holes},
Eur. Phys. J. C \textbf{80},  822 (2020).

\bibitem{Semi11}
I.~Semiz,
{ Dyonic Kerr-Newman black holes, complex scalar field and cosmic censorship},
Gen.\ Rel.\ Grav.\  {\bf 43}, 833 (2011).


\bibitem{DuSe13}
K.~D\"uztas and I.~Semiz,
{ Cosmic censorship, black holes and integer-spin test fields},
Phys.\ Rev.\ D {\bf 88}, 064043 (2013).


\bibitem{SeDu15}
I.~Semiz and K.~D\"uztas,
{ Weak cosmic censorship, superradiance and quantum particle creation},
Phys.\ Rev.\ D {\bf 92}, 104021 (2015).


\bibitem{Duzt15}
K.~D\"uztas,
{ Stability of event horizons against neutrino flux: the classical picture},
Class.\ Quant.\ Grav.\  {\bf 32}, 075003 (2015).


\bibitem{Gwak18}
B.~Gwak,
{ Weak cosmic censorship conjecture in Kerr-(Anti-)de Sitter black hole with scalar field},
JHEP {\bf 1809}, 081 (2018).

\bibitem{Gwak19a}
B.~Gwak,
{ Weak cosmic censorship with pressure and volume in charged Anti-de Sitter black hole under charged scalar field},
JCAP {\bf 1908}, 016 (2019).

\bibitem{Gwak19b}
B.~Gwak,
{ Weak cosmic censorship in Kerr-Sen black hole under charged scalar field},
JCAP \textbf{03}, 058 (2020).

\bibitem{Chen18}
D.~Chen,
{ Weak cosmic censorship conjecture in BTZ black holes with scalar fields},
Chin.\ Phys.\ C\ \textbf{44}, 015101 (2020).


\bibitem{YCWWL20}
S.~J.~Yang, J.~Chen, J.~J.~Wan, S.~W.~Wei, and Y.~X.~Liu,
{ Weak cosmic censorship conjecture for a Kerr-Taub-NUT black hole with a test scalar field and particle},
Phys.\ Rev.\ D \textbf{101}, 064048 (2020).



\bibitem{YWCYW20}
S.~J.~Yang, J.~J.~Wan, J.~Chen, J.~Yang, and Y.~Q.~Wang,
{ Weak cosmic censorship conjecture for the novel $4D$ charged Einstein-Gauss-Bonnet black hole with test scalar field and particle},
Eur.\ Phys.\ J.\ C {\bf 80}, 937 (2020).



\bibitem{LiWL19}
B.~Liang, S.~W.~Wei, and Y.~X.~Liu,
{ Weak cosmic censorship conjecture in Kerr black holes of modified gravity},
Mod.\ Phys.\ Lett.\ A {\bf 34}, 1950037 (2019).


\bibitem{Gwak17}
B.~Gwak,
{ Cosmic censorship conjecture in Kerr-Sen black hole},
Phys.\ Rev.\ D {\bf 95}, 124050 (2017).

\bibitem{MaSi07}
G.~E.~A.~Matsas and A.~R.~R.~da Silva,
{ Overspinning a nearly extreme charged black hole via a quantum tunneling process},
Phys.\ Rev.\ Lett.\  {\bf 99}, 181301 (2007).

\bibitem{Hod08}
S.~Hod,
{ Weak cosmic censorship: as strong as ever},
Phys.\ Rev.\ Lett.\  {\bf 100}, 121101 (2008).


\bibitem{RiSa08}
M.~Richartz and A.~Saa,
{ Overspinning a nearly extreme black hole and the weak cosmic censorship conjecture},
Phys.\ Rev.\ D {\bf 78}, 081503 (2008).


\bibitem{MRSSV09}
G.~E.~A.~Matsas, M.~Richartz, A.~Saa, A.~R.~R.~da Silva, and D.~A.~T.~Vanzella,
{ Can quantum mechanics fool the cosmic censor?},
Phys.\ Rev.\ D {\bf 79}, 101502 (2009).


\bibitem{RiSa11}
M.~Richartz and A.~Saa,
{ Challenging the weak cosmic censorship conjecture with charged quantum particles},
Phys.\ Rev.\ D {\bf 84}, 104021 (2011).

\bibitem{WLFY12}
S.~W.~Wei, Y.~X.~Liu, C.~E.~Fu, and K.~Yang,
{ Strong field limit analysis of gravitational lensing in Kerr-Taub-NUT spacetime},
JCAP {\bf 1210}, 053 (2012).


\bibitem{LCDJ11}
C.~Liu, S.~Chen, C.~Ding, and J.~Jing,
{ Particle acceleration on the background of the Kerr-Taub-NUT spacetime},
Phys.\ Lett.\ B\ \textbf{701}, 285 (2011).



\bibitem{JDLi19}
J.~Jiang, B.~Deng, and X.~W.~Li,
{ Holographic complexity of charged Taub-NUT-AdS black holes},
Phys.\ Rev.\ D\  {\bf 100}, 066007 (2019).


\bibitem{KLPe20}
G.~Kalamakis, R.~G.~Leigh, and A.~C.~Petkou,
{ Aspects of holography of Taub-NUT-AdS4},
Phys.\ Rev.\ D\  {\bf 103},126012 (2021).





\bibitem{taub51}
A.~H. Taub,
{ Empty space-times admitting a three parameter group of
	motions},
Ann. of Math.  472 (1951).

\bibitem{NeTU63}
E.~Newman, L.~Tamburino, and T.~Unti,
{ Empty-space generalization of the
	Schwarzschild metric},
J. Math. Phys. \textbf{4}, 915 (1963).

\bibitem{Misn63}
C.~W. Misner,
{{ The flatter regions of Newman, Unti, and Tamburino's
		generalized Schwarzschild space}},
J. Math. Phys. \textbf{4},  924 (1963).

\bibitem{Hawk73}
S.~W. Hawking and G.~F.~R. Ellis,
ph{The large scale structure of space-time}, vol.~1. Cambridge university press, 1973.

\bibitem{Haji71}
P.~Hajicek,
{{ Causality in non-Hausdorff space-times}},
Commun.\ Math.\ Phys.\ \textbf{21}, 75 (1971).

\bibitem{MaRu05}
V.~Manko and E.~Ruiz,
{{ Physical interpretation of the nut family of solutions}},
Class.\ Quant.\ Grav.\ \textbf{22}, 3555 (2005).


\bibitem{CGGu16}
G.~Cl\'ement, D.~Gal'tsov, and M.~Guenouche,
{ NUT wormholes},
Phys.\ Rev.\ D\  {\bf 93}, 024048 (2016).

\bibitem{CGGu15}
G.~Cl\'ement, D.~Gal'tsov, and M.~Guenouche,
{ Rehabilitating space-times with NUTs},
Phys.\ Lett.\ B\ \textbf{750}, 591 (2015).

\bibitem{CGGu18}
G.~Cl\'ement and M.~Guenouche,
{ Motion of charged particles in a NUTty Einstein-Maxwell spacetime and causality violation},
Gen.\ Rel.\ Grav.\ \textbf{50}, 60 (2018).




\bibitem{Hawk1998a}
S.~W. Hawking and C.~J. Hunter,
{{ Gravitational entropy and global structure}},
Phys.\ Rev.\ D\ \textbf{59}, 044025 (1999).


\bibitem{Hawk98b}
S.~W. Hawking, C.~J. Hunter, and D.~N. Page,
{{ NUT charge, anti-de Sitter space and entropy}},
Phys.\ Rev.\ D\ \textbf{59},  044033 (1999).


\bibitem{CEJM99}
A.~Chamblin, R.~Emparan, C.~V. Johnson, and R.~C. Myers,
{{ Large N phases, gravitational instantons and the nuts and bolts of AdS holography}},
Phys.\ Rev.\ D\ \textbf{59}, 064010 (1999).

\bibitem{EmJM99}
R.~Emparan, C.~V. Johnson, and R.~C. Myers,
{{ Surface terms as counterterms in the AdS/CFT correspondence}},
Phys.\ Rev.\ D\ \textbf{60}, 104001 (1999).

\bibitem{Mann99}
R.~B. Mann,
{{ Misner string entropy}},
Phys.\ Rev.\ D\ \textbf{60}, 104047 (1999).

\bibitem{Mann00}
R.~B. Mann,
{{ Entropy of rotating Misner string space-times}},
Phys.\ Rev.\ D\ \textbf{61}, 084013 (2000).

\bibitem{John14a}
C.~V. Johnson,
{{ Thermodynamic volumes for AdS-Taub--NUT and AdS-Taub-Bolt}},
Class.\ Quant.\ Grav.\ \textbf{31}, 235003 (2014).

\bibitem{John14b}
C.~V. Johnson,
{{ The extended thermodynamic phase structure of Taub--NUT and Taub-Bolt}},
Class.\ Quant.\ Grav.\ \textbf{31}, 225005 (2014).

\bibitem{GaMa00}
D.~Garfinkle and R.~B. Mann,
{{ Generalized entropy and Noether charge}},
Class.\ Quant.\ Grav.\ \textbf{17}, 3317 (2000).





\bibitem{CDLV19}
R.~Carballo-Rubio, F.~Di Filippo, S.~Liberati and M.~Visser,
{ Geodesically complete black holes},
Phys.\ Rev.\ D\ \textbf{101}, 084047 (2020).

\bibitem{ApGK17}
M.~Appels, R.~Gregory and D.~Kubiznak,
{ Black Hole Thermodynamics with Conical Defects},
JHEP \textbf{05}, 116 (2017).



\bibitem{BGHK19}
A.~B.~Bordo, F.~Gray, R.~A.~Hennigar, and D.~Kubiz\v{n}\'{a}k,
{ Misner gravitational charges and variable string strengths},
Class.\ Quant.\ Grav.\ \textbf{36}, 194001 (2019).



\bibitem{RMCh20}
M.~Rahman, S.~Mitra, and S.~Chakraborty,
{ Strong cosmic censorship conjecture with NUT charge and conformal coupling},
Class.\ Quant.\ Grav.\ \textbf{37}, 195004 (2020).

\bibitem{BCCa06}
E.~Berti, V.~Cardoso, and M.~Casals,
{ Eigenvalues and eigenfunctions of spin-weighted spheroidal harmonics in four and higher dimensions},
Phys.\ Rev.\ D \textbf{73}, 024013 (2006).




\bibitem{BCPa15}
R.~Brito, V.~Cardoso, and P.~Pani,
{ Superradiance: energy extraction, black-hole bombs and implications for astrophysics and particle physics},
Lect.\ Notes\ Phys.\ \textbf{906}, 1 (2015).

\bibitem{Hawk72}
S.~W.~Hawking,
{ Black holes in general relativity},
Commun.\ Math.\ Phys.\ \textbf{25}, 152 (1972).

\bibitem{CCFKNNB18}
M.~Cabero, C.~D.~Capano, O.~Fischer-Birnholtz, B.~Krishnan, A.~B.~Nielsen, A.~H.~Nitz, and C.~M.~Biwer,
{ Observational tests of the black hole area increase law},
Phys.\ Rev.\ D\ \textbf{97}, 124069 (2018).


\end{thebibliography}
\end{document}